\pdfoutput=1
\documentclass[runningheads]{llncs}

\usepackage{url}

\usepackage{amsmath}
\usepackage{amssymb}

\usepackage{cellspace}

\usepackage{pgfplots}

\usepackage{mathtools, nccmath}

\usepackage{tikz}
\usetikzlibrary{shapes}

\usepackage{algorithm}
\usepackage{algpseudocode}

\tikzset{
circ/.style={draw,circle,minimum height=3em},
}

\begin{document}
\setcounter{MaxMatrixCols}{50}

\title{On de Bruijn Rings and \\Families of Almost Perfect Maps}

\author{Peer Stelldinger\inst{1}}
\authorrunning{P. Stelldinger}
\institute{Department of Computer Science, HAW Hamburg, Germany \and
Hamburg University of Technology, Germany\\
\email{peer.stelldinger@haw-hamburg.de}\\
}
\maketitle              

\begin{abstract}
De Bruijn tori, or perfect maps, are two-dimensional periodic arrays of letters from a finite alphabet, where each possible pattern of shape $(m,n)$ appears exactly once in a single period. While the existence of certain de Bruijn tori, such as square tori with odd 
$m=n\in\{3,5,7\}$ and even alphabet sizes, remains unresolved, sub-perfect maps are often sufficient in applications like positional coding. These maps capture a large number of patterns, with each appearing at most once. While previous methods for generating such sub-perfect maps cover only a fraction of the possible patterns, we present a construction method for generating almost perfect maps for arbitrary pattern shapes and arbitrary non-prime alphabet sizes, including the above mentioned square tori with odd $m=n\in\{3,5,7\}$ as long that the alphabet size is non-prime. This is achieved through the introduction of de Bruijn rings, a minimal-height sub-perfect map and a formalization of the concept of families of almost perfect maps. The generated sub-perfect maps are easily decodable which makes them perfectly suitable for positional coding applications.
\end{abstract}


\section{Introduction}
De Bruijn tori, also called perfect maps, are two-dimensional cyclic matrices with their entries being drawn from some alphabet, such that every pattern of a given rectangular shape occurs exactly once within one period.

One of the most prominent use cases of de Bruijn tori is in the spatial coding context, e.g.\ for robot localization based on some optical ground pattern \cite{scheinerman2001,schuesselbauer2021}.

In such a use case, it is not required that every pattern occurs exactly once. It is sufficient, that every pattern occurs at most once. Such maps are called sub-perfect maps \cite{mitchell1994}.
For optical robot localization, a spatial code based on some $(m,n)$-pattern of letters from an alphabet of size $k$ should have the following properties:
\begin{enumerate}
    \item every local $(m,n)$-pattern occurs at most once.
    \item Almost every possible $(m,n)$-pattern is contained in the map.
    \item The code can efficiently be generated for different $m,n,k$.
    \item The decoding is efficient.
\end{enumerate}


The first two properties point to sub-perfect maps. Nevertheless, even in spatial coding, it is advantageous when the majority if not all patterns occur in the code. However, it is still unknown if certain types of de Bruijn tori exist, e.g.\ no square shaped de Bruijn torus with pattern shape $(n,n)$ is known for $n\in\{3,5,7\}$ and even alphabet size \cite{hurlbert1993}. For such cases, it is of interest to be able to generate sub-perfect maps which are not far away from perfect maps in that they contain almost every possible local pattern. 

For these applications, we provide a new efficient construction method for sub-perfect maps on arbitrary non-prime alphabet sizes including square shaped maps for any square pattern shape $(n,n)$ (including $n\in\{3,5,7\}$). Moreover, we prove, that these maps are almost perfect. This means for increasing $m,n$, the percentage of not occurring patterns tends to zero. 
The so-constructed sub-perfect maps are significantly larger than other known sub-perfect maps \cite{denes1990,makarov2019}.

While previous work either lacks efficient decoding methods \cite{hurlbert1995}, uses only a small subset of possible local patterns \cite{makarov2019} or use non-rectangular patterns \cite{bruckstein2012,horan2016}, the presented construction method is, to our knowledge, the first one giving an efficient decoding method for rectangular patterns on almost perfect maps.


\section{Basic Definitions and Related Work for the 1D Case}

Let an \emph{alphabet} be a totally ordered, nonempty finite set $\Sigma$ with its elements being called \emph{letters}. A \emph{word} is a finite sequence of letters $[a_1,a_2,\ldots,a_m]$ with $a_i\in\Sigma, 1\leq i\leq m$. A word is a \emph{Lyndon word}, if it is strictly less than any of the rotations $[a_i,\ldots,a_m,a_1,\ldots,a_{i-1}]$ $(2\leq i<m)$ with respect to the induced lexicographic order. With $lexmin(w)$ being the lexicographically smallest rotation of a word $w$, we call a word $w$ \emph{aperiodic}, if $lexmin(w)$ is a Lyndon word, otherwise $w$ is \emph{periodic}. The equivalence classes of words $w$ modulo cyclic rotation are called \emph{necklaces} with $lexmin(w)$ being a natural choice of representatives of these classes. Then the representatives of aperiodic necklaces are the Lyndon words.

For $|\Sigma|=k$, there are $k^m$ words of length $m$ in total. The number of Lyndon words is given by the necklace polynomial $\mathrm{M}(k,m)$ \cite{Moreau1872} (also known as Witt's formula \cite{lothaire1997}), as defined by
\[
    \mathrm{M}(k,m)=\frac1m\sum_{d|m}\mu\left(\frac md\right)k^d
\]
where $\mu$ is the classic M\"obius function \cite{Mobius1832}
\[
    \mu(n)=
        \begin{cases}
            1, & \text{if }n=1\\
            (-1)^i, & \text{if $n$ is the product of $i$ different primes}\\
            0, & \text{otherwise.}\\
        \end{cases}
\]
For $m$ prime this simplifies to $\mathrm{M}(k,m)=\frac1m(k^m-k)$. Note, that $m\cdot \mathrm{M}(k,m)$, which is the number of aperiodic words of a given length $m$, quickly tends to $k^m$ for increasing $m$. In other words, the majority of words of size $m$ are aperiodic and fall into equivalence classes of size $m$ under rotation \cite{Riedel2021}. As Riedel \cite{Riedel2021} only describes the asymptotic behaviour, we give an explicit (although not very strict) lower bound:

\begin{lemma}\label{lem:lowerbound}
Let $k,m\geq2$ be natural numbers. Then $\mathrm{M}(k,m)$ is smaller than $\frac1m k^m$ and larger than $\frac1m\left(k^m-k^{\lfloor m/2\rfloor+1}\right)$.
\end{lemma}
\begin{proof}
It is obvious that $\mathrm{M}(k,m)<\frac1m k^m$ since there are $k^m$ words of length $m$, each aperiodic necklace is an equivalence class of $m$ words and there are periodic necklaces for any $m\geq 2$. For the lower bound we have:
\begin{eqnarray*}
\mathrm{M}(k,m)&=&\frac1m\sum_{d|m}\mu\left(\frac md\right)k^d\\
&=&\frac1m\left(k^m+\sum_{d|m,\,\,d\neq m}\mu\left(\frac md\right)k^d\right)\\
&\geq&\frac1m\left(k^m-\sum_{d|m,\,\,d\neq m}k^d\right)\\
&\geq&\frac1m\left(k^m-\sum_{i=1}^{\lfloor m/2\rfloor} k^i\right)\\
&=&\frac1m\left(k^m-\frac{k\left(k^{\lfloor m/2\rfloor}-1\right)}{k-1}\right)\\
&>&\frac1m\left(k^m-\frac{k\left(k^{\lfloor m/2\rfloor}\right)}{k-1}\right)\\
&>&\frac1m\left(k^m-k^{\lfloor m/2\rfloor+1}\right)\\
\end{eqnarray*}
\end{proof}

Since the probability of a random word of length $m$ given an alphabet of size $k\geq 2$ being periodic is given by $1-m \mathrm{M}(k,m)/k^m$, it follows corollary \ref{cor:prop}:
\begin{corollary}\label{cor:prop}
    Given an alphabet $\Sigma$ with $\Sigma=|k|\geq 2$, the probability of a random word of length $m$ being periodic does not exceed $k^{1-\lceil m/2\rceil}$.
\end{corollary}

An \emph{$(M;\;m)_k$-de Bruijn sequence} is a cyclic sequence of letters from an alphabet $\Sigma$ of size $k$ in which every possible word of length $m$ occurs exactly once. Such a sequence always exist and is of length $M=k^m$ \cite{debruijn1946}. 
%

A well-known method to generate de Bruijn sequences is the construction of an Eulerian cycle or of a Hamiltonian cycle in a de Bruijn digraph \cite{klein2013}: 
Given an alphabet $\Sigma$ with $|\Sigma|=k$, the $m$-dimensional \emph{de Bruijn digraph} is defined by the vertex set $V=\Sigma^m$ of words of length $m$ and the edge set $E=\{(v,w)\mid v=[v_1,v_2,\ldots,v_m]\in V;\, w=[v_2,\ldots,v_m,v_{m+1}]\in V\}$. 
Any such graph is Hamiltonian and each Hamiltonian cycle defines a $(k^m;\;m)_k$-de Bruijn sequence, given by $[a_1,a_2,a_3,...,a_{k^m}]$ where each sub-sequence $[a_i,a_{i+1},...,a_{i+m}]$ of the cyclic sequence is the $i$th vertex of the Hamiltonian cycle $(1\leq i\leq k^m)$. 
Moreover, the $(m-1)$-dimensional de Bruijn digraph is Eulerian and each Euler cycle defines a $(k^m;\;m)_k$-de Bruijn sequence, given by $[a_1,a_2,a_3,...,a_{k^m}]$ where each length-$(m-1)$-subword $[a_i,a_{i+1},...,a_{i+m-1}]$ of the cyclic sequence is the $i$th vertex of the Eulerian cycle.

\section{Basic Definitions and Related Work for the 2D Case}

Given an alphabet $\Sigma$, a \emph{pattern} of shape $(m,n)$ is a finite array of letters 
\[
P=\begin{bmatrix}
    p_{1,1} & \cdots  & p_{1,n} \\
    \vdots & \ddots & \vdots \\
    p_{m,1} & \cdots  & p_{m,n}
\end{bmatrix}
\]
with $p_{i,j}\in\Sigma$, $1\leq i\leq m$ and $1\leq j\leq n$. A pattern is a \emph{row-Lyndon pattern}, if the word $[[p_{1,1},\ldots,p_{1,n}],\ldots[p_{m,1},\ldots,p_{m,n}]]$ is a Lyndon word given the alphabet $\Sigma^n$. Further, we define $lexmin(P)$ as the rotation of rows
\[
\begin{bmatrix}
    p_{i,1} & \cdots  & p_{i,n} \\
    \vdots & \ddots & \vdots \\
    p_{m,1} & \cdots  & p_{m,n} \\
    p_{1,1} & \cdots  & p_{1,n}\\
    \vdots & \ddots & \vdots \\
    p_{i-1,1} & \cdots  & p_{i-1,n},
\end{bmatrix}
\]
such that the corresponding word $[[p_{i,1},\ldots,p_{i,n}],\ldots,[p_{i-1,1},\ldots,p_{i-1,n}]]$ is a lexicographically smallest one. We call
a pattern \emph{row-aperiodic} if $lexmin(w)$ is a row-Lyndon pattern, otherwise it is \emph{row-periodic}.

A cyclic two-dimensional array of shape $(M,N)$ of letters from an alphabet $\Sigma$ of size $|\Sigma|=k$, is called an \emph{$(M,N;\; m,n)_k$-sub-perfect map}, if every $(m,n)$-pattern $p\in\Sigma^{m,n}$ occurs at most once. It is called an \emph{$(M,N;\; m,n)_k$-perfect map} or an \emph{$(M,N;\; m,n)_k$-de Bruijn torus} if every $(m,n)$-pattern occurs exactly once. $(M,N;\; m,n)_k$ is called the \emph{type} of the (sub-)perfect map.

It is well known \cite{paterson1994,hurlbert1993}, that (omitting the trivial cases $M=m=1$ and/or $N=n=1$) every $(M,N;\; m,n)_k$-de Bruijn torus must fulfill 
\begin{equation} \label{eqn:toruscond}
 M>m\text{, }N>n\text{ and }MN=k^{mn}\text{.}
\end{equation}

For $k=2$ it has been shown constructively, that these necessary conditions are also sufficient \cite{paterson1994}.
For $k>2$, the question is not that simple. However, there are a lot of methods known for the construction of de Bruijn tori.

E.g.\ Cook proves that for every $m, n$ and $k$ (except $n=2$ if $k$ even), there exists a $(k^m,k^{mn-m};\; m,n)_k$-de Bruijn torus \cite{cook1988}.

As shown by Hurlbert and Isaak \cite{hurlbert1993}, for $k$ even and $n\not\in\{3,5,7,9\}$, for $k=2$ and for $k$ odd there exists a $(k^{n^2/2},k^{n^2/2};\;n,n)_k$-de Bruijn torus if and only if $n$ is even or $k$ is a perfect square. They conjecture that there are also such square shaped de Bruijn tori for $k$ even and $n\in\{3,5,7,9\}$ iff $k$ is a perfect square, but this is still not proven.

Some previous work addresses the construction of sub-perfect maps. Notably, Makarov and Yashunsky\cite{makarov2019} describe a construction method for easily decodable $(M,N;\; m,n)_k$-sub-perfect maps, where a periodic base-pattern is combined with a sparse periodic pattern of markers by suitably increasing the base-pattern alphabet of size $d$ to some $k>d$. However, as the markers are only sparsely set, only a subset (at most $d^{mn-1}k$) of the $k^{mn}$ possible sub-patterns is used. Thus, the resulting size of the sub-perfect maps is significantly below the possible maximum. 

In the following we show on how to construct sub-perfect maps for the cases where no perfect maps are known and for other cases as well. Similarly to to the construction method of Makarov and Yashunsky, we combine a repeated base-pattern with a second one to increase the size of the total pattern while maintaining easy decodability. However, we use a second base-pattern of similar size instead of sparse markers and thus get a total pattern of almost maximal size given some alphabet $k$. As base-patterns we introduce de Bruijn rings:

\section{De Bruijn Rings}
When replacing the necessary condition $M>m$ in equation \ref{eqn:toruscond}  by $M=m$, one gets a sub-perfect map of minimal height. We denote such thin but long sub-perfect maps as de Bruijn rings. 
%

\begin{definition}[De Bruijn ring]\label{def:ring}
A $(m,N;\; m,n)_k$-sub-perfect map is called an \emph{$(m,n)_k$-de Bruijn ring} if there does not exist any $(m,N';\; m,n)_k$-sub-perfect map with $N'>N$.
\end{definition}

For $m=1$, a $(m,n)_k$-de Bruijn ring simplifies to a one-dimensional $(k^n;\;n)_k$-de Bruijn sequence. Therefore, we assume for the rest of the paper, that $m>1$. For similar reasons we also omit the trivial cases $n=1$ and $k=1$.
As an example, a $(2,2)_2$-de Bruijn ring is given by \begin{equation} \label{eqn:ring22}
\begin{bmatrix} 0 & 0 & 1 & 1 & 0 & 0\\ 0 & 1 & 0 & 1 & 1 & 1\end{bmatrix}\text{.}\end{equation}
This is a sub-perfect map which contains every possible binary $(2,2)$ pattern except of 
\[\begin{bmatrix} 0 & 0 \\ 0 & 0 \end{bmatrix}\text{, }
\begin{bmatrix} 0 & 1 \\ 0 & 1\end{bmatrix}\text{, }
\begin{bmatrix} 1 & 0 \\ 1 & 0\end{bmatrix}\text{ and } \begin{bmatrix} 1 & 1 \\ 1 & 1\end{bmatrix}\text{,}\]
since each of these patterns would occur twice within a cyclic map of height two.
It is obvious, that a $(m,N;\; m,n)_k$-de Bruijn Ring can only contain row-aperiodic $(m,n)$-patterns, since row-periodic patterns would occur more than once in the cyclic map. 
e.g.\ for $m=3, n=2$ and $k=2$, there are $2^6=64$ patterns in total, whereas the following $4$ patterns are the only row-periodic ones and thus can not occur in a $(3,2)_2$-de Bruijn Ring:
\[
\begin{bmatrix} 0 & 0 \\ 0 & 0 \\ 0 & 0\end{bmatrix}, 
\begin{bmatrix} 0 & 1 \\ 0 & 1 \\ 0 & 1\end{bmatrix}, 
\begin{bmatrix} 1 & 0 \\ 1 & 0 \\ 1 & 0\end{bmatrix}, 
\begin{bmatrix} 1 & 1 \\ 1 & 1 \\ 1 & 1\end{bmatrix}
\]
In case of $m=4, n=2$ and $k=2$, the following $16$ out of the possible $256$ patterns are row-periodic:
\[
\begin{bmatrix} 0 & 0 \\ 0 & 0 \\ 0 & 0 \\ 0 & 0\end{bmatrix}, 
\begin{bmatrix} 0 & 1 \\ 0 & 1 \\ 0 & 1 \\ 0 & 1\end{bmatrix}, 
\begin{bmatrix} 0 & 0 \\ 0 & 1 \\ 0 & 0 \\ 0 & 1\end{bmatrix}, 
\begin{bmatrix} 0 & 1 \\ 0 & 0 \\ 0 & 1 \\ 0 & 0\end{bmatrix}, 
\begin{bmatrix} 1 & 0 \\ 1 & 0 \\ 1 & 0 \\ 1 & 0\end{bmatrix}, 
\begin{bmatrix} 1 & 1 \\ 1 & 1 \\ 1 & 1 \\ 1 & 1\end{bmatrix}, 
\begin{bmatrix} 1 & 0 \\ 1 & 1 \\ 1 & 0 \\ 1 & 1\end{bmatrix}, 
\begin{bmatrix} 1 & 1 \\ 1 & 0 \\ 1 & 1 \\ 1 & 0\end{bmatrix},\]
\[
\begin{bmatrix} 0 & 0 \\ 1 & 0 \\ 0 & 0 \\ 1 & 0\end{bmatrix}, 
\begin{bmatrix} 0 & 1 \\ 1 & 1 \\ 0 & 1 \\ 1 & 1\end{bmatrix}, 
\begin{bmatrix} 0 & 0 \\ 1 & 1 \\ 0 & 0 \\ 1 & 1\end{bmatrix}, 
\begin{bmatrix} 0 & 1 \\ 1 & 0 \\ 0 & 1 \\ 1 & 0\end{bmatrix}, 
\begin{bmatrix} 1 & 0 \\ 0 & 0 \\ 1 & 0 \\ 0 & 0\end{bmatrix}, 
\begin{bmatrix} 1 & 1 \\ 0 & 1 \\ 1 & 1 \\ 0 & 1\end{bmatrix}, 
\begin{bmatrix} 1 & 0 \\ 0 & 1 \\ 1 & 0 \\ 0 & 1\end{bmatrix}, 
\begin{bmatrix} 1 & 1 \\ 0 & 0 \\ 1 & 1 \\ 0 & 0\end{bmatrix}.
\]

In general, given some alphabet size $k$, there are $k^{mn}$ patterns of shape $(m,n)$, of which $m\cdot \mathrm{M}(k^n,m)$ are row-aperiodic (each being one of the $m$ vertical rotations of a row-Lyndon pattern) and $k^{mn} - m\cdot \mathrm{M}(k^n,m)$ are row-periodic. The ratio of row-aperiodic patterns $ap(m,n,k):=(m\cdot \mathrm{M}(k^n,m))/k^{mn}$ quickly grows towards $1$ with increasing $m, n$ and/or $k$, as 
\[1\geq ap(m,n,k)> \frac{k^{mn}-k^{(\lfloor m/2\rfloor+1)n}}{k^{mn}},\]
and the limit of this fraction is $1$ for any of $m\to\infty$, $n\to\infty$ or $k\to\infty$. 
Table \ref{tab:percentages} in the appendix illustrates the ratio of row-aperiodic patterns for $2\leq m,n\leq 6$ and $2\leq k\leq 5$.
Further examples for de Bruijn rings of different sizes are given in the following:

\begin{eqnarray}
    &&(3,2)_2:\\\notag &&\begin{bmatrix} 
00000000100000101010\\
00010011010111011011\\
01101011111101011111
\end{bmatrix}\\\label{ring1}
&&(2,2)_3:\\\notag &&\begin{bmatrix} 
000121111110200021010112201000102021\\
011010201211212200200221221202221222
\end{bmatrix}\\\label{ring2}
&&(2,3)_2:\\\notag &&\begin{bmatrix} 
0000000100011001010111110110\\
0010111011001010011010011111
\end{bmatrix}\\\label{ring3}
&&(4,2)_2:\\\notag &&\begin{bmatrix} 
000010001000000100010101000111101000100010000001000010101110\\
000001100000010101111101111101101111010010000100011111010001\\
000100001001111001111010010000111101111010111010111101000101\\
011000110101101001101101010101111111011111101010110101111011
\end{bmatrix}\label{ring4}
\end{eqnarray}

Each of these $(m,n)_k$-de Bruijn rings contains every possible row-aperiodic $(m,n)_k$-pattern. Thus, these sub-perfect maps are of maximal width. The following theorem \ref{thm:ring} states, that this is always the case. There always exists an $(m,\mathrm{M}(k^n,m);\; m,n)_k$-sub-perfect map.

\begin{theorem}\label{thm:ring}
For any alphabet size $k$ and any pattern size $(m,n)$, there exists an $(m,\mathrm{M}(k^n,m);\; m,n)_k$ sub-perfect map. This map is an $(m,n)_k$-de Bruijn ring.
\end{theorem}

It is obvious, that only the $m\cdot \mathrm{M}(k^n,m)$ row-aperiodic patterns of size $(m,n)$ can occur in a map of height $m$. Thus, a sub-perfect map of height $m$ has a width of at most $\mathrm{M}(k^n,m)$. In order to prove that such a map exists and that it can efficiently be constructed, we introduce $(m,n)_k$-ring graphs, which generalize the construction of de Bruijn sequences using Euler cycles to two dimensions. In contrast to the one-dimensional case, these ring graphs are not necessarily Hamiltonian.

\begin{definition}[Ring graph]\label{def:ringgraph}
A \emph{$(m,n)_k$-ring graph} is a labeled multidigraph $G=(V,E)$ with the vertices $V=\{lexmin(S)\mid S\in\Sigma^{m,n-1}\}$ and $E$ being the set of labeled edges $(P_1,P_2)$ with label $R$, such that 
there exist patterns $L,R\in\Sigma^{m,1}$, $C\in\Sigma^{m,n-2}$, such that
\begin{enumerate}
    \item  $P_1=[L,C]\in V$, $P_2=lexmin([C,R])\in V$, 
    \item $lexmin([L,C,R])$ is a row-Lyndon pattern of shape $(m,n)$ and 
    \item there is no lexicographically smaller $R'<R$ with $R'\in\Sigma^{m,1}$ and $lexmin([L,C,R']) = lexmin([L,C,R])$. 
\end{enumerate}
We set one separate edge (each with a different label) for every $R\in\Sigma^{m,1}$ between $P_1$ and $P_2$ if more than one are applicable.
\end{definition}

The nodes are representatives of the vertical rotation groups of patterns of shape $(m,n-1)$. Each node is the representative of a set of vertical rotations:
\[
\begin{bmatrix}
    p_{1,1} & \cdots  & p_{1,n-1} \\
    \vdots & \ddots & \vdots \\
    p_{m,1} & \cdots  & p_{m,n-1}
\end{bmatrix}
,
\begin{bmatrix}
    p_{2,1} & \cdots  & p_{2,n-1} \\
    \vdots & \ddots & \vdots \\
    p_{m,1} & \cdots  & p_{m,n-1} \\
    p_{1,1} & \cdots  & p_{1,n-1}
\end{bmatrix}
, \dots
\begin{bmatrix}
    p_{m,1} & \cdots  & p_{m,n-1} \\
    p_{1,1} & \cdots  & p_{1,n-1} \\
    \vdots & \ddots & \vdots \\
    p_{m-1,1} & \cdots  & p_{m-1,n-1}
\end{bmatrix}
\]
The number of elements in each vertical rotation group is a divisor of $m$ and depends on the vertical periodicity of the group elements.

Moreover, each edge stands for the vertical rotation group of a unique pattern of shape $(m,n)$ and one gets a group representative by concatenating the outgoing vertex label with the edge label.

While any vertex $P$ is equal to $lexmin(P)$, this is not necessarily true for the edge labels. A vertex will have an outgoing edge for any $R\in\Sigma^{m,1}$, which completes the vertex label to a unique row-aperiodic pattern. E.g.\ in case of $m=2$, $n=3$ and $k=2$, there are two edges from the vertex $\left[\substack{0\\1}\substack{1\\1}\right]$ to the vertex $\left[\substack{1\\1}\substack{0\\1}\right]$, with one edge being labeled as $\left[\substack{0\\1}\right]$ and the other as $\left[\substack{1\\0}\right]$. These edges represent the row-Lyndon patterns $\left[\substack{0\\1}\substack{1\\1}\substack{0\\1}\right]$ and $\left[\substack{0\\1}\substack{1\\1}\substack{1\\0}\right]$. However, due to the third condition, there is only one edge from the vertex $\left[\substack{0\\0}\substack{0\\0}\right]$ to the vertex $\left[\substack{0\\0}\substack{0\\1}\right]$ with label $\left[\substack{0\\1}\right]$, which represents the row-Lyndon pattern $\left[\substack{0\\0}\substack{0\\0}\substack{0\\1}\right]$. 
Figures \ref{fig:graph1} to \ref{fig:graph2}, as well as \ref{fig:graph3} to \ref{fig:graph5} show the ring graphs which correspond to the de Bruijn rings given above. The following lemmata describe some interesting properties of ring graphs, which we will use for proving theorem \ref{thm:ring}:

\begin{lemma}
Any $(m,n)_k$-ring graph has exactly $\mathrm{M}(k^n,m)$ edges. 
\end{lemma}

\begin{proof}
By definition of the edges of a ring graph, each row-Lyndon pattern $P$ of shape $(m,n)$ corresponds to exactly one edge of label $R$ from a vertex $V$ to some other vertex by $P=lexmin([V,R])$. This implies the lemma. 
\end{proof}

\begin{lemma}
Any $(m,n)_k$-ring graph is weakly connected. 
\end{lemma}

\begin{proof}
Let $0,1\in\Sigma$ be two different characters with $0<1$. Now let $V$ be an arbitrary vertex of the ring graph. Since $V=lexmin(V)$ and $R:=[0,...,0,1]^T$ is row-Lyndon, the concatenation $[V,R]$ is row-Lyndon as well and thus $V$ has an outgoing edge being labelled with $R$. This edge directs to a vertex $V'$, of which the number of rightmost columns being equal to $R$ is increased by one compared to $V$ itself if not all columns of $V$ are already equal to $R$. In that last case, the edge goes from $V$ to $V$ itself. It follows that there is a directed path from every vertex to the vertex $[R,R,...,R]$.
\end{proof}

\begin{lemma}
Any $(m,n)_k$-ring graph is Eulerian. 
\end{lemma}

\begin{proof}
Let $[L,C]$ be a vertex of a ring graph with $L\in\Sigma^{m,1}$ and $C\in\Sigma^{m,n-2}$. Further, let $R\in\Sigma^{m,1}$. Then $lexmin([L,C,R])$ is a row-Lyndon pattern iff $lexmin([R,L,C])$ is a row-Lyndon pattern, since the row-periodicity does not depend on the order of the columns. With $lexmin([L,C,R])$ being row-Lyndon, let $L'\in\Sigma^{m,1}$, $C'\in\Sigma^{m,n-2}$ and $R'\in\Sigma^{m,1}$ be uniquely defined by $lexmin([R,L,C])=lexmin([L',C',R'])$, $[L',C']=lexmin([L',C'])$, and $R'$ being the lexicographically smallest such choice if $[L',C']$ is row-periodic. It follows that there is an edge of label $R$ from $[L,C]$ to $lexmin([C,R])$ iff there is an edge of label $R'$ from $[L',C']$ to $lexmin([C',R'])=[L,C]$. This defines a one-to-one mapping of ingoing and outgoing edges of the arbitrarily chosen vertex $[L,C]$. Thus, each vertex of the ring graph has its in-degree equal to its out-degree, and due to the ring graph being weakly connected, each ring graph is Eulerian.
\end{proof}

Now we prove theorem \ref{thm:ring} by constructing $(m,n)_k$-de Bruijn rings based on Eulerian cycles of $(m,n)_k$-ring graphs:

\begin{proof} [Proof of Theorem \ref{thm:ring}.]
Let $C=(V_0,R_1,V_1,R_2,V_2,\dots,R_{\mathrm{M}(k^n,m)},V_{\mathrm{M}(k^n,m)}=V_0)$ be an Eulerian cycle of an $(m,n)_k$-ring graph which starts and ends in the vertex $V_0:=0^{m,n-1}$ with $0\in\Sigma$. Such a cycle can be constructed in linear time by using Hierholzers algorithm \cite{hierholzer1873}. We now build a sub-perfect map using algorithm \ref{alg:ring}.

\begin{algorithm}
\caption{Generation of a de Bruijn ring. }\label{alg:ring}
\begin{algorithmic}
\Require An Eulerian cycle $C=(V_0,R_1,V_1,R_2,V_2,\dots,R_{\mathrm{M}(k^n,m)},V_{\mathrm{M}(k^n,m)})$ of an $(m,n)_k$-ring graph using the alphabet $\Sigma$.
\Require $V_0=0^{m,n-1}$ for some $0\in\Sigma$.\\
\State $Ring \gets V_0$
\For  {$i=1$ \bf{to} $\mathrm{M}(k^n,m)-(n-1)$}
    \State $Ring \gets [Ring, R_i]$
    \State Rotate $Ring$ vertically, such that it ends with $V_i$
\EndFor \\
\Return $Ring$
\end{algorithmic}
\end{algorithm}

In every round of the for loop, algorithm \ref{alg:ring} adds a new column to $Ring$. For $i=1$, $Ring$ is set equal to $[V_0,R_1]$ with $R_1$ being an edge from $V_0$ to $V_1$. $[V_0,R_1]$ is a row-Lyndon pattern which ends with $V_1$. Now suppose, after $i-1$ rounds of the loop, $Ring$ ends with $V_{i-1}$ for some $2\leq i\leq \mathrm{M}(k^n,m)-(n-1)$. Then there exists an edge $R_i$ from $V_{i-1}$ to $V_i$ and $lexmin([V_{i-1},R_i])$ is row-Lyndon. Moreover, the last $n-1$ columns of $Ring$ are equal to $V_i$ up to vertical rotation. Thus, vertically rotating $Ring$ accordingly lets it end with $V_i$. It follows by induction, that each $n$-column subset of $Ring$ is a vertical rotation of a unique row-Lyndon pattern. After $\mathrm{M}(k^n,m)-(n-1)$ loops, $Ring$ is of length $\mathrm{M}(k^n,m)$ and thus contains every row-aperiodic $(m,n)_k$ pattern. Moreover it is a cyclic map, since the remaining $n-1$ edges in $C$ lead to $V_\mathrm{M}(k^n,m)=V_0$ and thus are labelled with $0^m$. Thus, $Ring$ is an $(m,\mathrm{M}(k^n,m);\; m,n)_k$ sub-perfect map. Since there cannot be a bigger sub-perfect map of height $m$, it is an $(m,n)_k$-de Bruijn ring.
\end{proof}
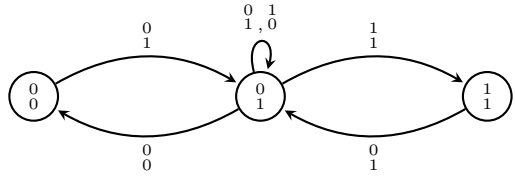
\begin{figure}[htbp]
    \centering
\begin{tikzpicture}[->,>=stealth,shorten >=1pt,auto,node distance=3cm,thick,node/.style={circle,draw}]
  \node[node] (00) {$\substack{0\\0}$};
  \node[node] (01) [right of=00] {$\substack{0\\1}$};
  \node[node] (11) [right of=01] {$\substack{1\\1}$};
  \path
    (00) edge [bend left] node {$\substack{0\\1}$} (01)
    (01) edge [loop above] node {$\substack{0\\1}$\,$\substack{\\\\,}$\,$\substack{1\\0}$} (01)
         edge [bend left] node {$\substack{1\\1}$} (11)     
         edge [bend left] node {$\substack{0\\0}$} (00)     
    (11) edge [bend left] node {$\substack{0\\1}$} (01);     
 \end{tikzpicture}
    \caption{The $(2,2)_2$-ring graph.}
    \label{fig:graph1}
\end{figure}
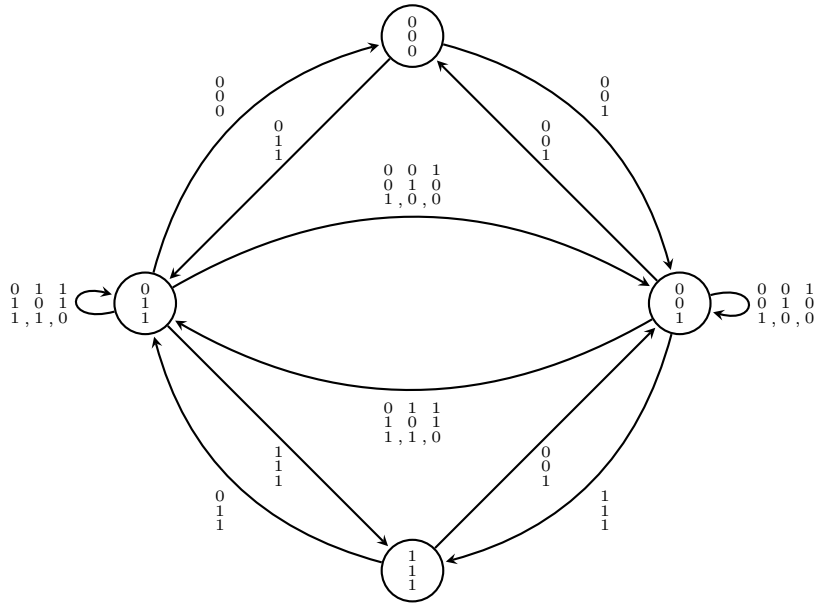
\begin{figure}[htbp]
    \centering
\begin{tikzpicture}[->,>=stealth,shorten >=1pt,auto,node distance=5cm,thick,node/.style={circle,draw}]
  \node[node] (000) {$\substack{0\\0\\0}$};
  \node[node] (001) [below right of=000] {$\substack{0\\0\\1}$};
  \node[node] (011) [below left  of=000] {$\substack{0\\1\\1}$};
  \node[node] (111) [below left  of=001] {$\substack{1\\1\\1}$};
  \path
    (000) edge [bend left] node {$\substack{0\\0\\1}$} (001)
          edge node [above] {$\substack{0\\1\\1}$} (011)
    (001) edge [loop right] node {$\substack{0\\0\\1}$\,$\substack{\\\\\\,}$\,$\substack{0\\1\\0}$\,$\substack{\\\\\\,}$\,$\substack{1\\0\\0}$} (001)
          edge [bend left] node {$\substack{1\\1\\1}$} (111)     
          edge [bend left] node {$\substack{0\\1\\1}$\,$\substack{\\\\\\,}$\,$\substack{1\\0\\1}$\,$\substack{\\\\\\,}$\,$\substack{1\\1\\0}$} (011)     
        edge node[above] {$\substack{0\\0\\1}$} (000)     
    (011) edge [bend left] node {$\substack{0\\0\\1}$\,$\substack{\\\\\\,}$\,$\substack{0\\1\\0}$\,$\substack{\\\\\\,}$\,$\substack{1\\0\\0}$} (001)
          edge node [below] {$\substack{1\\1\\1}$} (111)
          edge [loop left] node {$\substack{0\\1\\1}$\,$\substack{\\\\\\,}$\,$\substack{1\\0\\1}$\,$\substack{\\\\\\,}$\,$\substack{1\\1\\0}$} (011)
           edge [bend left] node {$\substack{0\\0\\0}$} (000)
    (111) edge [bend left] node {$\substack{0\\1\\1}$} (011)
          edge node [below] {$\substack{0\\0\\1}$} (001)
    ;     
\end{tikzpicture}
    \caption{The $(3,2)_2$-ring graph.}
    \label{fig:graph2}
\end{figure}

Note, that the generation of de Bruijn rings is both time- and space-efficient, as the construction of ring graphs, the construction of one of their Euler cycles using the Hierholzer algorithm and algorithm \ref{alg:ring} all have linear time and space complexity with respect to the number of aperiodic patterns.

In the following, we will use de Bruijn rings for constructing sub-perfect maps of more general sizes.

\section{From Rings to larger Sub-Perfect Maps}

Two sub-perfect maps with patterns of the same shape$(m,n)$ but not necessarily different alphabet sizes $k_1,k_2$  can be easily combined to a larger sub-perfect map with patterns of the same shape but larger alphabet size $k_1k_2$:

\begin{lemma}\label{lem:simplecombination}
Let $A_1$ be an $(M_1,N_1;\;m,n)_{k_1}$-sub-perfect map and $A_2$ be an $(M_2,N_2;\;m,n)_{k_2}$-sub-perfect map. Then there exists 
an $\left(M,N;\; m,n\right)_{k_1k_2}$-sub-perfect map $A$ with \[M=\mathrm{lcm}(M_1, M_2) \text{ and }N=\mathrm{lcm}(N_1,N_2).\]
\end{lemma}

\begin{proof}
    Let $\Sigma_1$ and  $\Sigma_2$ be the alphabets of $A_1$ and $A_2$, respectively. We now simply lay the cyclic maps onto each other by combining $\Sigma_1$ and  $\Sigma_2$ to the product alphabet $\Sigma_1\times\Sigma_2$, which has $k_1k_2$ elements. This new map has a vertical period length of $(\mathrm{lcm}\left(M_1,M_2\right)$ and a horizontal period length of $\mathrm{lcm}\left(N_1,N_2\right)$.
\end{proof}

Note, that if both $A_1$ and $A_2$ are perfect maps, the combined map $A$ is a perfect map as well, iff $M_1$ and $M_2$ are coprime and $N_1$ and $N_2$ are coprime.

Lemma \ref{lem:simplecombination} can easily be applied to de Bruijn rings. E.g.\ when rotating one of two de Bruijn rings by $90^\circ$, it follows corollary \ref{lem:simpleRings}.

\begin{corollary}\label{lem:simpleRings}
Given two alphabets $\Sigma_1$ and  $\Sigma_2$ with $|\Sigma_1|=k_1$ and  $|\Sigma_2|=k_2$, there exists an $\left(M,N;\; m,n\right)_{k_1k_2}$-sub-perfect map with \[M=\mathrm{lcm}\left(m,\mathrm{M}(k_2^m,n)\right)\text{ and }N=\mathrm{lcm}\left(n,\mathrm{M}(k_1^n,m)\right).\]
\end{corollary}

\begin{proof}
According to theorem \ref{thm:ring}, there exists an $(m,n)_{k_1}$-de Bruijn ring $A_1$ and an $(n,m)_{k_2}$-de Bruijn ring $A_2$. By rotating $A_2$ by $90^\circ$, we get an $(\mathrm{M}(k_2^m,n),n;\; m,n)_k$-sub-perfect map $A_2'$. 
By stacking $A_1$ and $A_2'$ onto each other and combining the alphabets to $\Sigma=\Sigma_1\times\Sigma_2$, we get an  $(\mathrm{lcm}\left(m,\mathrm{M}k_2^m,n)\right),$\\ $\mathrm{lcm}\left(n,\mathrm{M}(k_1^n,m)\right);\; m,n)_{k_1k_2}$-sub-perfect map.
\end{proof}

For example, using the $(2,2)_2$-de Bruijn ring given in (\ref{eqn:ring22}) for both $A_1$ and $A_2$, and when interpreting the tuples of $\Sigma=\Sigma_1\times\Sigma_2$ as two-digit binary numbers and writing them as decimals, one gets the following $(6,6;\; 2,2)_4$-sub-perfect map:
\[
2
\begin{bmatrix} 
0  &   0  &   1  &   1  &   0  &   0\\
0  &   1  &   0  &   1  &   1  &   1\\
0  &   0  &   1  &   1  &   0  &   0\\
0  &   1  &   0  &   1  &   1  &   1\\
0  &   0  &   1  &   1  &   0  &   0\\
0  &   1  &   0  &   1  &   1  &   1
\end{bmatrix}
+
\begin{bmatrix} 
0  &   0  &   0  &   0  &   0  &   0\\
1  &   0  &   1  &   0  &   1  &   0\\
0  &   1  &   0  &   1  &   0  &   1\\
1  &   1  &   1  &   1  &   1  &   1\\
1  &   0  &   1  &   0  &   1  &   0\\
1  &   0  &   1  &   0  &   1  &   0
\end{bmatrix}\
=
\begin{bmatrix} 
0  &   0  &   2  &   2  &   0  &   0\\
1  &   2  &   1  &   2  &   3  &   2\\
0  &   1  &   2  &   3  &   0  &   1\\
1  &   3  &   1  &   3  &   3  &   3\\
1  &   0  &   3  &   2  &   1  &   0\\
1  &   2  &   1  &   2  &   3  &   2
\end{bmatrix}\]

However, since the with $6$ and the height $2$ are not coprime, the resulting map is far away from containing almost all patterns. In fact, it contains only $36$ of $256$ possible $(2,2)$-patterns of a $4$-letter alphabet. A simple trick is to remove the fifth column of the de Bruijn ring and use this sub-perfect map as $A_1$ and $A_2$. Now, the width and height are coprime and the resulting combined sub-perfect map has size $(10,10)$:

\[\begin{bmatrix} 
0 & 0 & 2 & 2 & 0 & 0 & 0 & 2 & 2 & 0\\
1 & 2 & 1 & 2 & 3 & 0 & 3 & 0 & 3 & 2\\
0 & 1 & 2 & 3 & 0 & 1 & 0 & 3 & 2 & 1\\
1 & 3 & 1 & 3 & 3 & 1 & 3 & 1 & 3 & 3\\
1 & 0 & 3 & 2 & 1 & 0 & 1 & 2 & 3 & 0\\
0 & 2 & 0 & 2 & 2 & 0 & 2 & 0 & 2 & 2\\
1 & 0 & 3 & 2 & 1 & 0 & 1 & 2 & 3 & 0\\
0 & 3 & 0 & 3 & 2 & 1 & 2 & 1 & 2 & 3\\
1 & 1 & 3 & 3 & 1 & 1 & 1 & 3 & 3 & 1\\
1 & 2 & 1 & 2 & 3 & 0 & 3 & 0 & 3 & 2
\end{bmatrix}\]

In this case, the number of covered patterns increases to almost three times the original number, but is still less than half the number of all possible patterns. However, as we show, the trick of removing columns to ensure coprime dimensions of $A_1$ and $A_2$ can be used in general, and in case of larger rings, the percentage of covered patterns quickly grows near $100\%$ with increasing $m$, $n$. E.g. when using the $(3,2)_2$-de Bruijn ring (\ref{ring1}) as $A_1$ and the $(2,3)_2$-de Bruijn ring (\ref{ring3}) as $A_2$, one gets a $(84, 20;\; 3,2)_4$-sub-perfect map without using the trick and a $(84, 38;\; 3,2)_2$-sub-perfect map by removing the second column from $A_1$. This map covers $3192$ of $4096$ patterns. Both sub-perfect maps are given in figure \ref{fig:maps} in the appendix.

In general, if $m$ and $\mathrm{M}(k_2^m,n)$ are not coprime or if $n$ and $\mathrm{M}(k_1^n,m)$ are not coprime, then the resulting sub-perfect map contains only a fraction of all possible $(m,n)$-patterns over $\Sigma$ (at most half of them). 

However, we are interested in families of almost perfect maps, where the maps contain almost every pattern.

\begin{definition}[Almost perfect map]
Let $k>2$ and $\mathcal{M}$ be a set of sub-perfect maps for an alphabet of size $k$, such that for any tuple $m',n'\geq 1$ there exists a $(M,N;\;m,n)_k$-sub-perfect map in $\mathcal{M}$ with $m>m'$ and $n>n'$.
A sequence $(A_i)_{i\in\mathbb{N}}$ of sub-perfect maps for an alphabet of size $k$ is called an \emph{increasing sequence} if for any elements $A_i$ of type $(M_i,N_i;\;m_i,n_i)_{k}$ 
and $A_j$ of type $(M_j,N_j;\;m_j,n_j)_{k}$ with $j>i$ it follows $m_j>m_i$ and $n_j>n_i$. 
We call a set of sub-perfect maps $\mathcal{M}$ a \emph{family of almost perfect maps}, if for any increasing sequence $(A_i)_{i\in\mathbb{N}}$ with $A_i$ being of type $(M_i,N_i;\;m_i,n_i)_k$ it follows
\begin{equation*}
\lim_{i \rightarrow \infty} \frac{M_iN_i}{k^{m_in_i}} = 1\text{.}
\end{equation*}
We call $\mathcal{M}$ a \emph{family of perfect maps}, if $M_iN_i=k^{m_in_i}$ for any element of type $(M_i,N_i;\;m_i,n_i)_k$. 
\end{definition}

Note, that the sub-perfect maps given by Makarov and Yashunsky \cite{makarov2019} are not a family of almost perfect maps: given some alphabet size $k$ where $0<d<k$ letters are used for the base code and $k-d>1$ letters as markers, the percentage of patterns included is given by $\frac{d^{mn-1}k}{k^{mn}}\leq \frac{(k-1)^{mn-1}k}{k^{mn}}=(\frac{k-1}{k})^{mn-1}$, which is $0$ for $m,n\rightarrow\infty$.

Theorem \ref{thm:ring} directly implies that $(m,n)_k$-de Bruijn rings are a family of almost perfect maps for any $k$, while the construction method given in lemma \ref{lem:simplecombination} does not give a family of almost perfect maps since $m$ and $\mathrm{M}(k_2^m,n)$, as well as $n$ and $\mathrm{M}(k_1^n,m)$ are not necessarily coprime. However, the trick being used above can be used for arbitrary de Bruijn rings, such that stacking them leads to families of almost perfect maps. Note, that in the following we use the Iverson bracket notation $[condition]$ \cite{iverson1962,knuth1992} which is $1$ if the $condition$ is true and $0$ otherwise.

\begin{theorem}\label{thm:bigRings}
Given two alphabets $\Sigma_1$ and  $\Sigma_2$ with $|\Sigma_1|=k_1\geq2$ and  
$|\Sigma_2|=k_2\geq2$, and given a pattern of shape $(m,n)$ with $|m-n|\leq 2$, there exists an $\left(M,N;\; m,n\right)_{k_1k_2}$-sub-perfect map with 
\begin{eqnarray*}
M&=&m\cdot\left(\mathrm{M}(k_2^m,n)-\left(\left|m'-1\right|\cdot [m'<m-1]\right)\right)\\
&\geq&m\cdot\left(\mathrm{M}(k_2^m,n)-\max(1,m-3)\right)
\end{eqnarray*}
and
\begin{eqnarray*}
N&=&n\cdot\left(\mathrm{M}(k_1^n,m)-\left(\left|n'-1\right|\cdot [n'<n-1]\right)\right)\\
&\geq&n\cdot\left(\mathrm{M}(k_1^n,m)-\max(1,n-3)\right)\text{,}
\end{eqnarray*}
where \[m':=\mathrm{M}(k_2^m,n) \bmod m\] and \[n':=\mathrm{M}(k_1^n,m) \bmod n\text{.}\]
\end{theorem}

\begin{proof}
By definition of $m'$ it follows that $\mathrm{M}(k_2^m,n)-m'$ is a multiple of $m$. Thus, $\mathrm{M}(k_2^m,n)-m'\pm1$ and $m$ are coprime. Now let 
\[m'':=|m'-1|\cdot [m'<m-1]\text{.}\]
Then $m'$ and $m''$ either differ by one, or $m'=m-1$ and $m''=0$. Thus $\mathrm{M}(k_2^m,n)-m''$ and $m$ are coprime. Since $0\leq m''\leq \max(1,m-3)$, it follows 
\begin{eqnarray*}
M&:=&\mathrm{lcm}(m,\mathrm{M}(k_2^m,n)-m'')\\
&=&m\cdot(\mathrm{M}(k_2^m,n)-m'')\\
&\geq& m\cdot(\mathrm{M}(k_2^m,n)-\max(1,m-3))\text{.}
\end{eqnarray*}
Analogously it follows 
\begin{eqnarray*}
N&:=&\mathrm{lcm}(n,\mathrm{M}(k_1^n,m)-n'')\\
&=&n\cdot(\mathrm{M}(k_1^n,m)-n'')\\
&\geq& n\cdot(\mathrm{M}(k_1^n,m)-\max(1,n-3))
\end{eqnarray*}
with \[n'':=|n'-1|\cdot [n'<n-1]\text{.}\]

It remains to be shown, that in case of $n''>0$ and/or $m''>0$ there exist sub-perfect maps of types 
\[\left(m,\mathrm{M}(k_1^n,m)-m'';\;m,n\right)_{k_1}\text{ and}\]
\[\left(n,\mathrm{M}(k_2^m,n)-n'';\;n,m\right)_{k_2}\text{.}\]

Given some alphabet $\Sigma_1$, $|\Sigma_1|=k_1$, let $A$ be an $(m,n)_{k_1}$-de Bruijn ring, i.e.\ an $\left(m,\mathrm{M}(k_1^n,m);\;m,n\right)_{k_1}$-sub-perfect map. Now let $0,1\in\Sigma_1$. Then each of the $(m,n)$-patterns with the first $i$ rows ($1\leq i<m$) being equal to $0$ and the last $m-i$ rows being equal to $1$ is contained in $A$. Removing one column of one such pattern from $A$ removes this pattern and its vertical rotations from $A$ but still results in a sub-perfect map. Since there are $m-1$ such patterns, we can construct an $\left(m,\mathrm{M}(k_1^n,m)-i;\;m,n\right)_{k_1}$-sub-perfect map for any $1\leq i\leq m-1$ which includes $i:=n''\leq\max(1,n-3)\leq m-1$ since $|m-n|\leq 2$.

Analogously, it follows the existence of an $\left(n,\mathrm{M}(k_2^m,n)-j;\;n,m\right)_{k_2}$-sub-perfect map for any $1\leq j\leq n-1$ which includes $j:=m''\leq \max(1,m-3)\leq n-1$ since $|m-n|\leq 2$.
\end{proof}

\begin{theorem}\label{thm:bigRings2}
The set of sub-perfect maps given in theorem \ref{thm:bigRings} for $k=k_1k_2$ with $k_1,k_2\geq 2$ is a family of almost perfect maps.
\end{theorem}

\begin{proof}
Let  $(A_i)_{i\in\mathbb{N}}$ be an increasing sequence of such $\left(M_i,N_i;\; m_i,n_i\right)_{k}$-sub-perfect maps. Then for $k=k_1k_2$ it follows
\begin{eqnarray*}
\frac{M_iN_i}{k^{m_i n_i}} &\geq& \frac{m_i\left(\mathrm{M}(k_2^{m_i},n_i)-m_i+3\right)\cdot n_i\left(\mathrm{M}(k_1^{n_i},m_i)-n_i+3\right)}{(k_1k_2)^{m_i n_i}} \\
&=& \frac{\left(m_i\mathrm{M}(k_2^{m_i},n_i)-m_i^2+3m_i\right)\cdot \left(n_i\mathrm{M}(k_1^{n_i},m_i)-n_i^2+3n_i\right)}{k_2^{m_i n_i}k_{1,i}^{m_i n_i}} \\
&>&\frac{k_2^{m_in_i}-k_2^{(\lfloor m_i/2\rfloor+1)n_i}-m_i^2+3m_i}{k_2^{m_i n_i}}
\cdot\frac{k_1^{n_im_i}-k_1^{(\lfloor n_i/2\rfloor+1)m_i}-n_i^2+3n_i}{k_1^{m_i n_i}}\\
\end{eqnarray*}
For $k_1,k_2\geq 2$ and $m_i,n_i\rightarrow\infty$, the last term reduces to $\frac{k_2^{m_in_i}}{k_2^{m_in_i}}\cdot \frac{k_1^{m_in_i}}{k_1^{m_in_i}}$, thus $\lim_{i \rightarrow \infty} \frac{M_iN_i}{k^{m_in_i}} = 1$.
\end{proof}

%
%
In case of square shaped patterns $m=n$ and perfect square alphabet sizes $k_1=k_2$, we get square shaped maps:

\begin{corollary}\label{cor:sqrRings}
Given an alphabet $\Sigma$ with $|\Sigma|=k\geq 2$, there exists for any $n\geq 2$ a $\left(N,N;\; n,n\right)_{k^2}$-sub-perfect map with 
\begin{eqnarray*}
N&=&n\cdot\left(\mathrm{M}(k^n,n)-\left(\left|n'-1\right|\cdot [n'<n-1]\right)\right)\\
&\geq&n\cdot\left(\mathrm{M}(k^n,n)-\max(1,n-3)\right)\\
&>&k^{n^2}-k^{(\lfloor n/2\rfloor+1)n}-\max(n,n^2-3n)\text{,}
\end{eqnarray*}
with \[n'=\mathrm{M}(k^n,n) \bmod n\]
and this set of sub-perfect maps is a family of almost perfect maps.
\end{corollary}


\begin{theorem}\label{thm:sqrRingsPrime}
Given any alphabet $\Sigma$ with $|\Sigma|=k\geq 2$ and any prime $p$, there exists a $\left(N,N;\; p,p\right)_{k^2}$-sub-perfect map with 
$N=p\cdot(\mathrm{M}(k^p,p)-1)=k^{p^2}-k^p-p$.
\end{theorem}

\begin{proof}
    Since $p$ is prime, it follows $\mathrm{M}(k^p,p)=\frac{k^{p^2}-k^p}p$. If the 
    nominator $k^{p^2}-k^p$ is a multiple of $p^2$, it follows, that $\mathrm{M}(k^p,p)$ is a multiple of $p$ and thus $\mathrm{M}(k^p,p)-1$ is coprime to $p$. Then, according to corollary \ref{cor:sqrRings} there exists a $\left(N,N;\; p,p\right)_{k^2}$-sub-perfect map with $N=p\cdot(\mathrm{M}(k^p,p)-1)=k^{p^2}-k^p-p$.
    Thus, we only need to show that $k^{p^2}\equiv k^p \pmod{p^2}$, or equivalently the $p$-adic valuation $\nu_p(k^{p^2}-k^p)$ is at least $2$.

    Depending on $p$, the lifting-the-exponent (LTE) lemma \cite{parvadi2011} implies the following:
    \begin{itemize}
    \item{$p\mid k$:} In this case obviously $p^2\mid k^p$ and $p^2\mid k^{p^2}$, which implies $0\equiv k^{p^2}\equiv k^p \pmod{p^2}$.
    \item{$p\nmid k$, $p=2$:} Due to the LTE lemma it follows $\nu_2(k^{p^2}-k^p)=\nu_2(k^2-k)+\nu_2(k^2+k)+\nu_2(2)-1=\nu_2(k^2-k)+\nu_2(k^2+k)$, which is at least $2$ since $k^2\pm k$ is always even. 
    \item{$p\nmid k$, $p$ odd:} This implies $p\nmid k^p$. In this case, the LTE lemma gives $\nu_p(k^{p^2}-k^p)=\nu_p(k^p-k)+\nu_p(p)=\nu_p(k^p-k)+1$, which is at least $2$ due to Fermat's little theorem. 
\end{itemize}
\end{proof}

\section{Application for Optical Localization}

Theorems \ref{thm:bigRings} and \ref{thm:sqrRingsPrime} allow to generate sub-perfect maps which are perfectly suitable for optical localization tasks, since:
\begin{enumerate}
    \item every local $(m,n)$-pattern occurs at most once.
    \item Almost every possible $(m,n)$-pattern is contained in the map.
    \item The code can efficiently be generated for arbitrary $m,n,k\geq 2$ with $k$ being not prime (such that it can be decomposed into some $k_1$ and $k_2$).
    \item The decoding is efficient, as one only needs to store a $(m,n)_k$-de Bruijn ring, which reduces the number of entries to the square root of the whole pattern. 
\end{enumerate}

As an example, from theorem \ref{thm:sqrRingsPrime} follows that there is a $(501,501;\;3,3)_4$-sub-perfect map, which covers more than $95.7\%$ of all possible $(3,3)$ patterns of an alphabet of size $2^2=4$. Such a map requires just to build a $(3,3)_2$-de Bruijn ring, which is a matrix of size $(3,168)$ and which can efficiently be constructed. For example, with every tile being a square of 1cm side length (and the 2 bits being represented as one of four possible colors), the whole pattern would be of size $501\times 501$ cm with every local $3\times 3$ cm sub-pattern being unique.

\begin{figure}[ht]
    \centering
    \begin{tikzpicture}
        \begin{axis}[
            title={Percentage of covered sub-patterns},
            xlabel={Square pattern width $n$},
            ylabel={Coverage $N^2/\tilde{N}^2$ in $\%$},
            legend style={
                at={(1,0)}, 
                anchor=south east, 
                legend columns=1, 
                font=\small,
                fill=none, 
                draw=none, 
                column sep=1ex
            },
            grid=both,
            width=0.8\textwidth,
            xtick={2,3,4,5,6}, 
            xticklabel style={anchor=north}, 
            enlargelimits=true        ]
            \addplot[
                color=blue,
                mark=square,
                mark options={fill=blue}
            ]
            coordinates {
                (2,39.0625) (3,95.7492828369141) (4,99.2081169039011) (5,99.9997794629359) (6,99.9992253099205)
            };
            \addlegendentry{$k=2$}

            \addplot[
                color=red,
                mark=triangle,
                mark options={fill=red}
            ]
            coordinates {
                (2,74.6837372351776) (3,99.6954007251795) (4,99.9695005833771) (5,99.9999999414603) (6,99.9999994830579)
            };
            \addlegendentry{$k=3$}

            \addplot[
                color=green,
                mark=o,
                mark options={fill=green}
            ]
            coordinates {
                (2,86.431884765625) (3,99.9488895889954) (4,99.9969480792089) (5,99.9999999998172) (6,99.9999999970889)
            };
            \addlegendentry{$k=4$}

            \addplot[
                color=purple,
                mark=x,
                mark options={fill=purple}
            ]
            coordinates {
                (2,91.54662399999999) (3,99.9868932294967) (4,99.9994879954125) (5,99.9999999999979) (6,99.9999999999476)
            };
            \addlegendentry{$k=5$}
        \end{axis}
    \end{tikzpicture}
    \caption{Size comparison of square shaped sub-perfect maps according to theorem \ref{thm:sqrRingsPrime} and (not guaranteed to exist) de Bruijn tori of the same type. When $N$ is the side length of a sub-perfect map, $\tilde{N}$ denotes the side length of the corresponding de Bruijn torus. The plot shows the percentage of $N^2/\tilde{N}^2$, i.e. the ratio of covered to all $(n,n)_{k^2}$-patterns for  different values of $k$.}
    \label{fig:bigmap}
\end{figure}
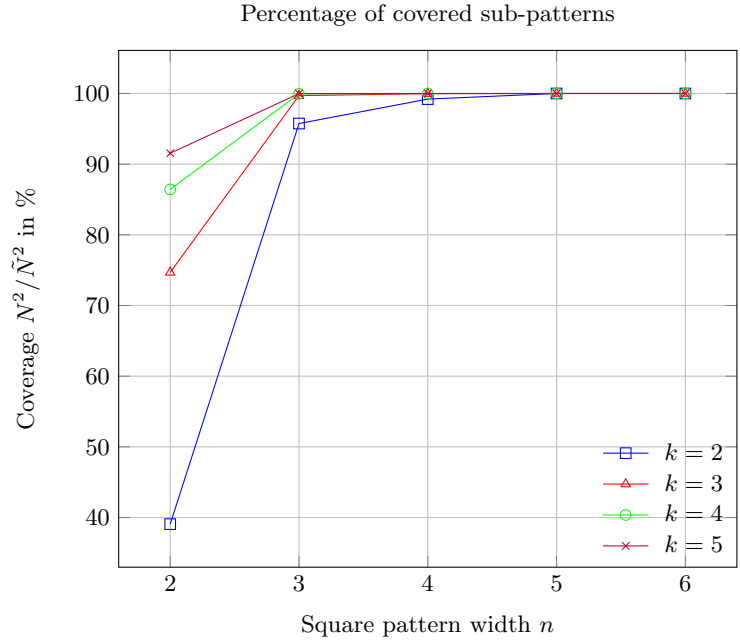

Analogously, a $(5,5)_2$-de Bruijn ring, which is of size $(5,6710880)$, can be used to construct a $(33 554 395, 33 554 395;\;5,5)_4$-sub-perfect map, which covers more than $99.9997\%$ of all $(5,5)$ patterns of an alphabet of size $4$, and a $(7,7)_2$-de Bruijn ring of size $(7, 80 421 421 917 312)$ gives a $(562 949 953 421 177,$ $562 949 953 421 177;\;7,7)_4$-sub-perfect map covering more than $99.999999999\%$ of all patterns. Figure \ref{fig:bigmap} shows the approximative behaviour for further $k$ and $n$. The exact numbers are given in table \ref{tab:bigmap} in the appendix.

The efficiency of the decoding means, that given some local pattern it is easily be possible to derive its position in the whole map. For sub-perfect maps being constructed according to theorem \ref{thm:bigRings} and its specializations, this can be done as follows: 
\begin{enumerate}
    \item The local $(m,n)$-pattern of alphabet size $k=k_1k_2$ is separated into the two patterns of alphabet sizes $k_1$ and $k_2$.
    \item Two lookup tables of size $k_1^{mn}$ respectively $k_2^{mn}$ are used to derive the relative position of the pattern in the two given modified de Bruijn rings.
    \item The absolute position in the combined map is computed from these relative positions. 
\end{enumerate}
The time complexity of steps 1 to 3 is $O(mn)$, $O(1)$ (lookup table with direct memory access) and $O(1)$, respectively. In total, the time complexity of such a decoding algorithm is $O(mn)$ and thus linear with regard to the size of the sub-pattern. The space complexity is linear with regard to the size of the de Bruijn rings, since they define the size of the lookup table. As the combined size of the rings is bounded by twice the size of the larger ring, the space complexity is bounded by $O(\max(k_1,k_2)^{mn})$. Since the size $MN$ of the total pattern is of order $O(k_1^{mn}k_2^{mn})\geq O(\min(k_1,k_2)^{2mn})$, the space complexity of the decoding algorithm is of order $O(\sqrt{MN})$ as long as $M$ and $N$ differ only by a constant factor.

If space complexity needs to be further restricted at the cost of time complexity, one could generate the de Bruijn rings implicitly for every lookup (with choosing a fixed selection scheme for each next edge in the linear-time Hierholzer algorithm) which results in time complexity $O(\sqrt{MN})$ and space complexity $O(mn)$ when the ring graphs are not stored permanently.

\section{Conclusion and Future Work}

We have introduced de Bruijn rings as sub-perfect maps of minimal height. We showed that these can be constructed by generating Eulerian cycles in a graph, which generalizes a construction method of de Bruijn sequences to two dimensions. The time complexity of this construction is linear in the number of contained patterns and a de Bruijn ring contains every pattern which does not only consists of constant columns.

We further introduced the concept of a family of almost perfect maps and show, that de Bruijn rings are almost perfect in that sense, which practically means, that they contain almost all possible patterns. Moreover, we showed, how de Bruijn rings can be used to construct families of almost perfect maps of other rectangular shapes, including known square shaped cases for which perfect maps are not yet known. The construction allows a decoding scheme which is sublinear (with regard to the total number of patterns) in time and space complexity. 

The findings are especially of practical interest in the spacial coding context, since here a map does not need to be perfect, but it should be efficiently constructible and decodable. 

In future, the findings could easily be generalized to higher dimensions. Moreover, it would be interesting to combine the findings with the non-rectangular patterns proposed by Horan and Stevens \cite{horan2016}. Another future research direction could be to investigate the possibility to build families of almost perfect maps for prime alphabet sizes like e.g. the binary alphabet $k=2$. Moreover, especially in the spacial coding context, the question arises if it is possible to construct de Bruijn rings where every pattern is unique under rotation as well. Sub-perfect maps based on such rings would robots allow to self-locate even when they are rotated by $90^\circ$ or more.


\appendix
\newpage

\section{Additional Tables and Figures}

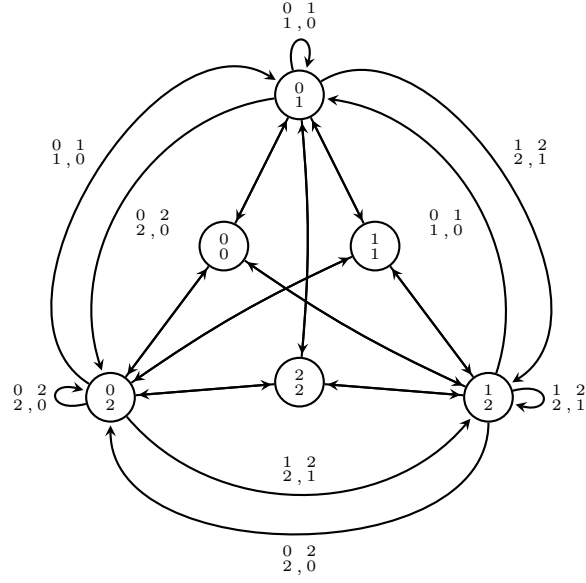
\begin{figure}[htbp]
    \centering
\begin{tikzpicture}[->,>=stealth,shorten >=1pt,auto,node distance=5cm,thick,node/.style={circle,draw}]
  \node[node] at (3,2.5) (00) {$\substack{0\\0}$};
  \node[node] at (4,4.5) (01) {$\substack{0\\1}$};
  \node[node] at (5,2.5) (11) {$\substack{1\\1}$};
  \node[node] at (1.5,0.5) (02) {$\substack{0\\2}$};
  \node[node] at (6.5,0.5) (12) {$\substack{1\\2}$};
  \node[node] at (4,0.7) (22) {$\substack{2\\2}$};
  \path
    (00) edge node [left]  {} (01)
         edge node [right] {} (02)
         edge [bend right=5] node {} (12)
    (11) edge node [left]  {} (12)
         edge node [right] {} (01)
         edge [bend right=5] node {} (02)
    (22) edge node [left]  {} (02)
         edge node [right] {} (12)
         edge [bend right=5] node {} (01)
    (01) edge [loop above] node {$\substack{0\\1}$\,$\substack{\\\\,}$\,$\substack{1\\0}$} (01)
         edge node [left] {} (11)     
         edge [bend left=5] node {} (22)     
         edge [bend right=50] node {$\substack{0\\2}$\,$\substack{\\\\,}$\,$\substack{2\\0}$} (02)     
         edge [bend left=90] node {$\substack{1\\2}$\,$\substack{\\\\,}$\,$\substack{2\\1}$} (12)     
         edge node [below] {} (00) 
    (12) edge [loop right] node {$\substack{1\\2}$\,$\substack{\\\\,}$\,$\substack{2\\1}$} (12)
         edge node [left]  {} (22)     
         edge [bend left=5] node {} (00) 
         edge [bend right=50] node {$\substack{0\\1}$\,$\substack{\\\\,}$\,$\substack{1\\0}$} (01)     
         edge [bend left=90] node {$\substack{0\\2}$\,$\substack{\\\\,}$\,$\substack{2\\0}$} (02)     
         edge node [below] {} (11) 
    (02) edge [loop left] node {$\substack{0\\2}$\,$\substack{\\\\,}$\,$\substack{2\\0}$} (02)
         edge node [left] {} (00)     
         edge [bend left=5] node {} (11)     
         edge [bend right=50] node {$\substack{1\\2}$\,$\substack{\\\\,}$\,$\substack{2\\1}$} (12)     
         edge [bend left=90] node {$\substack{0\\1}$\,$\substack{\\\\,}$\,$\substack{1\\0}$} (01)     
         edge node [below] {} (22); 
\end{tikzpicture}
    \caption{The $(2,2)_3$-ring graph. If no label is given at an edge, the label is equal to the target node label. A bidirectional edge represents two edges in opposing directions (with not necessarily equal labels).}
    \label{fig:graph3}
\end{figure}

\begin{figure}[htbp]
    \centering
\begin{tikzpicture}[->,>=stealth,shorten >=1pt,auto,node distance=4cm,thick,node/.style={circle,draw}]
  \node[node] at (0,3) (0000) {$\substack{00\\00}$};
  \node[node] at (2,2) (0010) {$\substack{00\\10}$};
  \node[node] at (2,4) (0001) {$\substack{00\\01}$};
  \node[node] at (5,0) (0101) {$\substack{01\\01}$};
  \node[node] at (5,2) (0110) {$\substack{01\\10}$};
  \node[node] at (5,4) (0011) {$\substack{00\\11}$};
  \node[node] at (5,6) (1010) {$\substack{10\\10}$};
  \node[node] at (8,2) (1011) {$\substack{10\\11}$};
  \node[node] at (8,4) (0111) {$\substack{01\\11}$};
  \node[node] at (10,3) (1111) {$\substack{11\\11}$};
  \path
    (0000) edge node [above]  {} (0001)
    (0001) edge node [above]  {} (0011)
           edge node [above]  {} (0110)
           edge [bend right] node {} (0010)     
           edge [bend left=40] node {} (0111)     
    (0010) edge node [above]  {} (0000)
           edge [bend right] node  {} (0101)
           edge node [right]  {$\substack{0\\1}$\,$\substack{\\\\,}$\,$\substack{1\\0}$} (0001)
    (1010) edge [bend right] node {} (0001)
    (0011) edge node [above]  {} (0111)
           edge node [above]  {} (0110)
           edge node [above]  {} (0010)
           edge [loop above] node {} (0011)
    (0110) edge node [above]  {} (0010)
           edge node [above]  {} (0011)
           edge node [above]  {} (0111)
           edge [loop below] node {} (0110)
    (0101) edge [bend right] node {} (1011)
    (0111) edge node [above]  {} (1111)
           edge [bend right] node  {} (1010)
           edge node [left]  {$\substack{0\\1}$\,$\substack{\\\\,}$\,$\substack{1\\0}$} (1011)
    (1011) edge node [above]  {} (0110)
           edge node [above]  {} (0011)
           edge [bend right] node {} (0111)     
           edge [bend left=40] node {} (0010)
    (1111) edge node [above]  {} (1011);
\end{tikzpicture}
    \caption{The $(2,3)_2$-ring graph. If no label is given at an edge, the label follows uniquely from the rules given in definition \ref{def:ringgraph}.}
    \label{fig:graph4}
\end{figure}
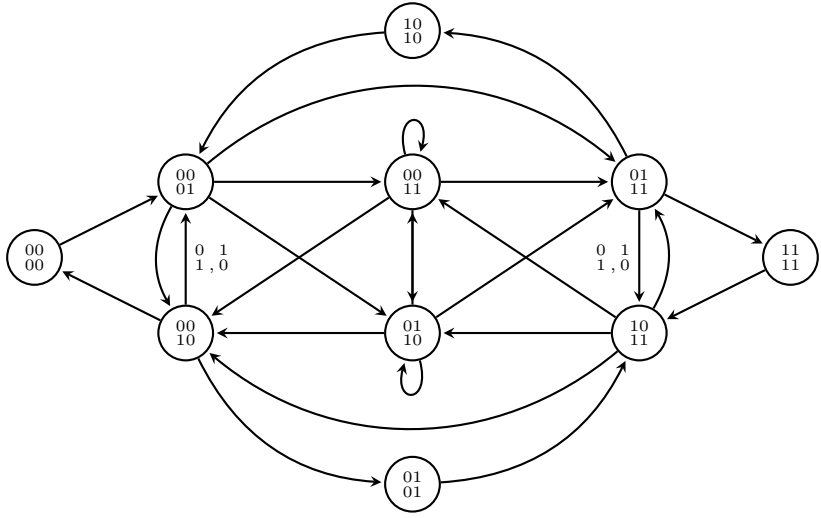

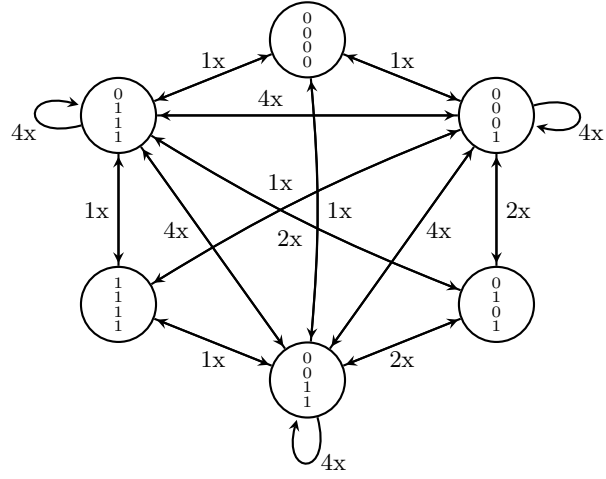
\begin{figure}[htbp]
    \centering
\begin{tikzpicture}[->,>=stealth,shorten >=1pt,auto,node distance=5cm,thick,node/.style={circle,draw}]
  \node[node] at (1.5,3.5) (0111) {$\substack{0\\1\\1\\1}$};
  \node[node] at (4,4.5) (0000) {$\substack{0\\0\\0\\0}$};
  \node[node] at (6.5,3.5) (0001) {$\substack{0\\0\\0\\1}$};
  \node[node] at (1.5,1) (1111) {$\substack{1\\1\\1\\1}$};
  \node[node] at (6.5,1) (0101) {$\substack{0\\1\\0\\1}$};
  \node[node] at (4,0) (0011) {$\substack{0\\0\\1\\1}$};
  \path
    (0111) edge [loop left] node [below] {4x\quad\quad} (0111)
           edge node [above]  {1x} (0000)
           edge node [] {} (1111)
           edge [bend right=5] node {} (0101)
           edge node [above] {4x\quad\quad\quad\quad} (0001)
           edge node [above] {4x\quad\quad\quad\quad} (0011)
    (0001) edge [loop right] node [below] {\quad4x} (0001)
           edge node []  {} (0101)
           edge node [] {} (0000)
           edge [bend right=5] node {} (1111)
           edge node [above] {\quad\quad\quad4x} (0011)
           edge node [] {} (0111)
    (0011) edge [loop below] node [right] {\,4x} (0011)
           edge node []  {} (1111)
           edge node [] {} (0101)
           edge [bend right=5] node {} (0000)
           edge node [] {} (0111)
           edge node [] {} (0001)
    (0000) edge node  [above] {1x}(0001)     
           edge [bend left=5] node {1x} (0011)     
           edge node {} (0111) 
    (0101) edge node [below]  {2x} (0011)     
           edge [bend left=5] node [below] {2x\quad\quad} (0111) 
           edge node [right] {2x} (0001) 
    (1111) edge node [left] {1x} (0111)     
           edge [bend left=5] node {1x} (0001)     
           edge node [below] {1x} (0011); 
\end{tikzpicture}
    \caption{The $(4,2)_2$-ring graph. For better readability, the edge labels (which are rotations of the vertex label the edge is pointing to) are omitted and only the number of labels per edge and direction is given.}
    \label{fig:graph5}
\end{figure}

\begin{figure}[htbp]
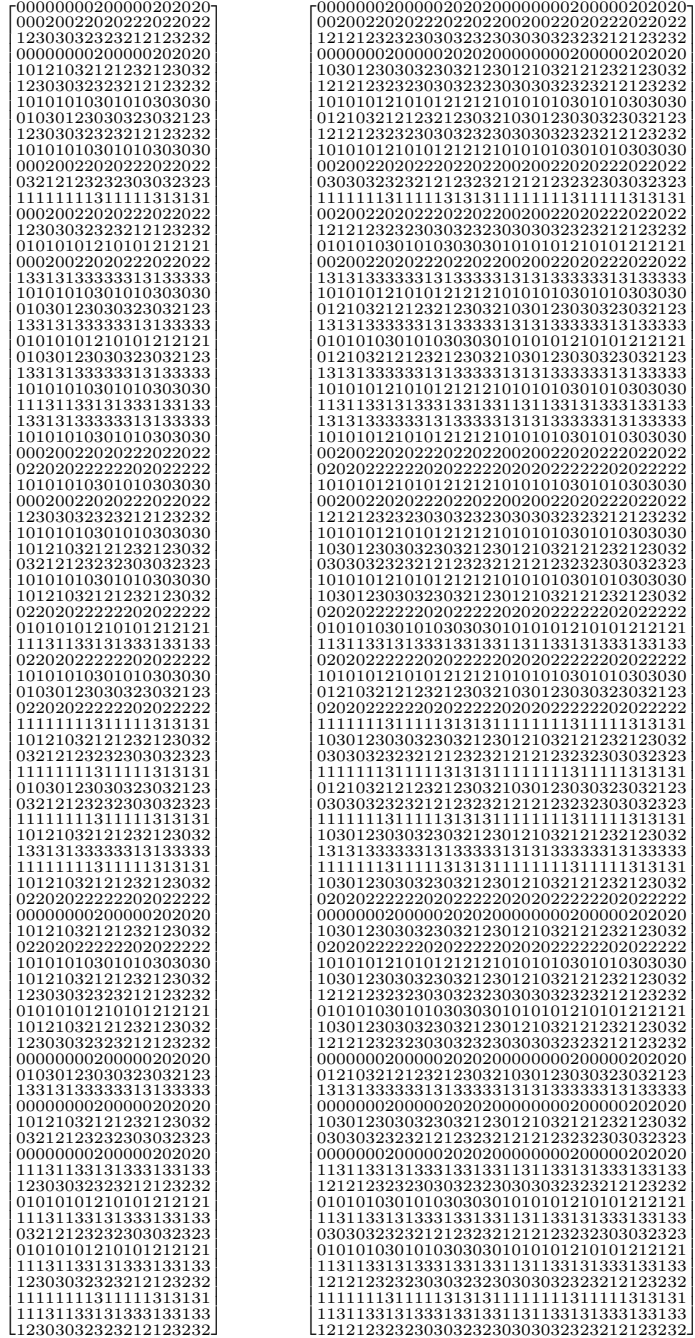

    \centering
{\fontsize{6pt}{6pt}\selectfont
\[\begin{bmatrix}
00000000200000202020\\
00020022020222022022\\
12303032323212123232\\
00000000200000202020\\
10121032121232123032\\
12303032323212123232\\
10101010301010303030\\
01030123030323032123\\
12303032323212123232\\
10101010301010303030\\
00020022020222022022\\
03212123232303032323\\
11111111311111313131\\
00020022020222022022\\
12303032323212123232\\
01010101210101212121\\
00020022020222022022\\
13313133333313133333\\
10101010301010303030\\
01030123030323032123\\
13313133333313133333\\
01010101210101212121\\
01030123030323032123\\
13313133333313133333\\
10101010301010303030\\
11131133131333133133\\
13313133333313133333\\
10101010301010303030\\
00020022020222022022\\
02202022222202022222\\
10101010301010303030\\
00020022020222022022\\
12303032323212123232\\
10101010301010303030\\
10121032121232123032\\
03212123232303032323\\
10101010301010303030\\
10121032121232123032\\
02202022222202022222\\
01010101210101212121\\
11131133131333133133\\
02202022222202022222\\
10101010301010303030\\
01030123030323032123\\
02202022222202022222\\
11111111311111313131\\
10121032121232123032\\
03212123232303032323\\
11111111311111313131\\
01030123030323032123\\
03212123232303032323\\
11111111311111313131\\
10121032121232123032\\
13313133333313133333\\
11111111311111313131\\
10121032121232123032\\
02202022222202022222\\
00000000200000202020\\
10121032121232123032\\
02202022222202022222\\
10101010301010303030\\
10121032121232123032\\
12303032323212123232\\
01010101210101212121\\
10121032121232123032\\
12303032323212123232\\
00000000200000202020\\
01030123030323032123\\
13313133333313133333\\
00000000200000202020\\
10121032121232123032\\
03212123232303032323\\
00000000200000202020\\
11131133131333133133\\
12303032323212123232\\
01010101210101212121\\
11131133131333133133\\
03212123232303032323\\
01010101210101212121\\
11131133131333133133\\
12303032323212123232\\
11111111311111313131\\
11131133131333133133\\
12303032323212123232
\end{bmatrix}\qquad\qquad
\begin{bmatrix}
00000002000002020200000000200000202020\\
00200220202220220220020022020222022022\\
12121232323030323230303032323212123232\\
00000002000002020200000000200000202020\\
10301230303230321230121032121232123032\\
12121232323030323230303032323212123232\\
10101012101012121210101010301010303030\\
01210321212321230321030123030323032123\\
12121232323030323230303032323212123232\\
10101012101012121210101010301010303030\\
00200220202220220220020022020222022022\\
03030323232121232321212123232303032323\\
11111113111113131311111111311111313131\\
00200220202220220220020022020222022022\\
12121232323030323230303032323212123232\\
01010103010103030301010101210101212121\\
00200220202220220220020022020222022022\\
13131333333131333331313133333313133333\\
10101012101012121210101010301010303030\\
01210321212321230321030123030323032123\\
13131333333131333331313133333313133333\\
01010103010103030301010101210101212121\\
01210321212321230321030123030323032123\\
13131333333131333331313133333313133333\\
10101012101012121210101010301010303030\\
11311331313331331331131133131333133133\\
13131333333131333331313133333313133333\\
10101012101012121210101010301010303030\\
00200220202220220220020022020222022022\\
02020222222020222220202022222202022222\\
10101012101012121210101010301010303030\\
00200220202220220220020022020222022022\\
12121232323030323230303032323212123232\\
10101012101012121210101010301010303030\\
10301230303230321230121032121232123032\\
03030323232121232321212123232303032323\\
10101012101012121210101010301010303030\\
10301230303230321230121032121232123032\\
02020222222020222220202022222202022222\\
01010103010103030301010101210101212121\\
11311331313331331331131133131333133133\\
02020222222020222220202022222202022222\\
10101012101012121210101010301010303030\\
01210321212321230321030123030323032123\\
02020222222020222220202022222202022222\\
11111113111113131311111111311111313131\\
10301230303230321230121032121232123032\\
03030323232121232321212123232303032323\\
11111113111113131311111111311111313131\\
01210321212321230321030123030323032123\\
03030323232121232321212123232303032323\\
11111113111113131311111111311111313131\\
10301230303230321230121032121232123032\\
13131333333131333331313133333313133333\\
11111113111113131311111111311111313131\\
10301230303230321230121032121232123032\\
02020222222020222220202022222202022222\\
00000002000002020200000000200000202020\\
10301230303230321230121032121232123032\\
02020222222020222220202022222202022222\\
10101012101012121210101010301010303030\\
10301230303230321230121032121232123032\\
12121232323030323230303032323212123232\\
01010103010103030301010101210101212121\\
10301230303230321230121032121232123032\\
12121232323030323230303032323212123232\\
00000002000002020200000000200000202020\\
01210321212321230321030123030323032123\\
13131333333131333331313133333313133333\\
00000002000002020200000000200000202020\\
10301230303230321230121032121232123032\\
03030323232121232321212123232303032323\\
00000002000002020200000000200000202020\\
11311331313331331331131133131333133133\\
12121232323030323230303032323212123232\\
01010103010103030301010101210101212121\\
11311331313331331331131133131333133133\\
03030323232121232321212123232303032323\\
01010103010103030301010101210101212121\\
11311331313331331331131133131333133133\\
12121232323030323230303032323212123232\\
11111113111113131311111111311111313131\\
11311331313331331331131133131333133133\\
12121232323030323230303032323212123232
\end{bmatrix}\]
}
    \caption{Sub-perfect maps generated by combining the de Bruijn rings (\ref{ring1}) and (\ref{ring3}) without (left) and with (right) making the combined widths and heights coprime.}
    \label{fig:maps}
\end{figure}

\begin{table}
     \centering
     \renewcommand{\arraystretch}{1.3}
     \begin{tabular}{c|c||c|c|c|c|}
          $m$&$n$& $k=2$& $k=3$& $k=4$& $k=5$\\
          \hline
          \hline
         $2$&$2$ & $\frac{12}{16}$ & $\frac{72}{81}$ & $\frac{240}{256}$ & $\frac{600}{625}$\\
          \hline
         $2$&$3$ & $\frac{56}{64}$ & $\frac{702}{729}$ & $\frac{4032}{4096}$ & $\frac{15500}{15625}$\\
          \hline
         $2$&$4$ & $\frac{240}{256}$ & $\frac{6480}{6561}$ & $\frac{65280}{65536}$ & $\frac{390000}{390625}$\\
          \hline
         $2$&$5$ & $\frac{992}{1024}$ & $\frac{58806}{59049}$ & $\frac{1047552}{1048576}$ & $\frac{9762500}{9765625}$\\
          \hline
         $2$&$6$ & $\frac{4032}{4096}$ & $\frac{530712}{531441}$ & $\frac{16773120}{16777216}$ & $\frac{244125000}{244140625}$\\
          \hline
          \hline
         $3$&$2$ & $\frac{60}{64}$ & $\frac{720}{729}$ & $\frac{4080}{4096}$ & $\frac{15600}{15625}$\\
          \hline
         $3$&$3$ & $\frac{504}{512}$ & $\frac{19656}{19683}$ & $\frac{262080}{262144}$ & $\frac{1953000}{1953125}$\\
          \hline
         $3$&$4$ & $\frac{4080}{4096}$ & $\frac{531360}{531441}$ & $\frac{16776960}{16777216}$ & $\frac{244140000}{244140625}$\\
          \hline
         $3$&$5$ & $\frac{32736}{32768}$ & $\frac{14348664}{14348907}$ & $\frac{1073740800}{1073741824}$ & $\frac{30517575000}{30517578125}$\\
          \hline
         $3$&$6$ & $\frac{262080}{262144}$ & $\frac{387419760}{387420489}$ & $\frac{68719472640}{68719476736}$ & $\frac{3814697250000}{3814697265625}$\\
          \hline
          \hline
         $4$&$2$ & $\frac{240}{256}$ & $\frac{6480}{6561}$ & $\frac{65280}{65536}$ & $\frac{390000}{390625}$\\
          \hline
         $4$&$3$ & $\frac{4032}{4096}$ & $\frac{530712}{531441}$ & $\frac{16773120}{16777216}$ & $\frac{244125000}{244140625}$\\
          \hline
         $4$&$4$ & $\frac{65280}{65536}$ & $\frac{43040160}{43046721}$ & $\frac{4294901760}{4294967296}$ & $\frac{152587500000}{152587890625}$\\
          \hline
         $4$&$5$ & $\frac{1047552}{1048576}$ & $\frac{3486725352}{3486784401}$ & $\frac{1099510579200}{1099511627776}$ & $\frac{95367421875000}{95367431640625}$\\
          \hline
         $4$&$6$ & $\frac{16773120}{16777216}$ & $\frac{282429005040}{282429536481}$ & $\frac{281474959933440}{281474976710656}$ & $\frac{59604644531250000}{59604644775390624}$\\
          \hline
          \hline
         $5$&$2$ & $\frac{1020}{1024}$ & $\frac{59040}{59049}$ & $\frac{1048560}{1048576}$ & $\frac{9765600}{9765625}$\\
          \hline
         $5$&$3$ & $\frac{32760}{32768}$ & $\frac{14348880}{14348907}$ & $\frac{1073741760}{1073741824}$ & $\frac{30517578000}{30517578125}$\\
          \hline
         $5$&$4$ & $\frac{1048560}{1048576}$ & $\frac{3486784320}{3486784401}$ & $\frac{1099511627520}{1099511627776}$ & $\frac{95367431640000}{95367431640625}$\\
          \hline
         $5$&$5$ & $\frac{33554400}{33554432}$ & $\frac{847288609200}{847288609443}$ & $\frac{1125899906841600}{1125899906842624}$ & $\frac{298023223876950016}{298023223876953152}$\\
          \hline
         $5$&$6$ & $\frac{1073741760}{1073741824}$ & $\frac{205891132093920}{205891132094649}$ & $\frac{1152921504606846976}{1152921504606846976}$ & $\frac{931322574615478534144}{931322574615478534144}$\\
          \hline
          \hline
         $6$&$2$ & $\frac{4020}{4096}$ & $\frac{530640}{531441}$ & $\frac{16772880}{16777216}$ & $\frac{244124400}{244140625}$\\
          \hline
         $6$&$3$ & $\frac{261576}{262144}$ & $\frac{387400104}{387420489}$ & $\frac{68719210560}{68719476736}$ & $\frac{3814695297000}{3814697265625}$\\
          \hline
         $6$&$4$ & $\frac{16772880}{16777216}$ & $\frac{282428998560}{282429536481}$ & $\frac{281474959868160}{281474976710656}$ & $\frac{59604644530860000}{59604644775390624}$\\
          \hline
         $6$&$5$ & $\frac{1073708064}{1073741824}$ & $\frac{205891117686936}{205891132094649}$ & $\frac{1152921503532057600}{1152921504606846976}$ & $\frac{931322574584951078912}{931322574615478534144}$\\
          \hline
         $6$&$6$ & $\frac{68719210560}{68719476736}$ & $\frac{150094634909047936}{150094635296999136}$ & $\frac{4722366482800908959744}{4722366482869645213696}$ & $\frac{14551915228363036345499648}{14551915228366852423942144}$\\
          \hline
          \hline
     \end{tabular}
     \caption{Ratio of row-aperiodic to all patterns of shape $(m,n)$ and alphabet size $k$.}
     \label{tab:percentages}
 \end{table}

\begin{table}
     \centering
     \renewcommand{\arraystretch}{1.15}
     \begin{tabular}{c|c||c|c|}
          $k$&$n$& $N/\tilde{N}$& $N^2/\tilde{N}^2$ in $\%$\\
          \hline
          \hline
         $2$&$2$ & $\frac{10}{16}$ & $39.062500000000000\%$\\
          \hline
         $2$&$3$ & $\frac{501}{512}$ & $95.749282836914063\%$\\
          \hline
         $2$&$4$ & $\frac{65276}{65536}$ & $99.208116903901100\%$\\
          \hline
         $2$&$5$ & $\frac{33554395}{33554432}$ & $99.999779462935919\%$\\
          \hline
         $2$&$6$ & $\frac{68719210554}{68719476736}$ & $99.999225309920504\%$\\
          \hline
          \hline
         $3$&$2$ & $\frac{70}{81}$ & $74.683737235177560\%$\\
          \hline
         $3$&$3$ & $\frac{19653}{19683}$ & $99.695400725179510\%$\\
          \hline
         $3$&$4$ & $\frac{43040156}{43046721}$ & $99.969500583377069\%$\\
          \hline
         $3$&$5$ & $\frac{847288609195}{847288609443}$ & $99.999999941460331\%$\\
          \hline
         $3$&$6$ & $\frac{150094634909047936}{150094635296999136}$ & $99.999999483057906\%$\\
          \hline
          \hline
         $4$&$2$ & $\frac{238}{256}$ & $86.431884765625000\%$\\
          \hline
         $4$&$3$ & $\frac{262077}{262144}$ & $99.948889588995371\%$\\
          \hline
         $4$&$4$ & $\frac{4294901756}{4294967296}$ & $99.996948079208892\%$\\
          \hline
         $4$&$5$ & $\frac{1125899906841595}{1125899906842624}$ & $99.999999999817220\%$\\
          \hline
         $4$&$6$ & $\frac{4722366482800908959744}{4722366482869645213696}$ & $99.999999997088906\%$\\
          \hline
          \hline
         $5$&$2$ & $\frac{598}{625}$ & $91.546623999999994\%$\\
          \hline
         $5$&$3$ & $\frac{1952997}{1953125}$ & $99.986893229496729\%$\\
          \hline
         $5$&$4$ & $\frac{152587499996}{152587890625}$ & $99.999487995412494\%$\\
          \hline
         $5$&$5$ & $\frac{298023223876950016}{298023223876953152}$ & $99.999999999997897\%$\\
          \hline
         $5$&$6$ & $\frac{14551915228363036345499648}{14551915228366852423942144}$ & $99.999999999947562\%$\\
          \hline
          \hline
     \end{tabular}

     \caption{Size comparison of square shaped sub-perfect maps according to theorem \ref{thm:sqrRingsPrime} and (not guaranteed to exist) de Bruijn tori of the same type. When $N$ is the side length of a sub-perfect map, $\tilde{N}$ denotes the side length of the corresponding de Bruijn torus. The second column gives the ratio of covered to all $(n,n)_{k^2}$-patterns.}
     \label{tab:bigmap}
 \end{table}




\end{document}